\documentclass[preprint,12pt]{elsarticle}

\usepackage{amssymb}
\usepackage{amsmath,amsfonts}
\usepackage{amsthm}
\usepackage{mathrsfs}
\usepackage{graphicx}
\usepackage{multirow}
\usepackage{xcolor}
\usepackage{textcomp}
\usepackage{booktabs}
\usepackage{hyperref}
\usepackage{gensymb}
\usepackage{siunitx}
\usepackage[T1]{fontenc}
\usepackage[utf8]{inputenc}

\journal{Radiation Physics and Chemistry}

\begin{document}

\begin{frontmatter}

\title{\textbf{High-Intense Gamma-Ray Emission from a Crystalline Undulator with Realistic Bending Profiles}}

\author[1,2]{R.~Negrello\corref{cor1}}
\ead{riccardo.negrello@unife.it}
\author[2]{L.~Malagutti}
\author[2]{L.~Bandiera\corref{cor2}}
\ead{laura.bandiera@fe.infn.it}
\author[2]{N.~Canale}
\author[1,2]{F.~Cescato}
\author[1,2]{P.~Fedeli}
\author[1,2]{V.~Guidi}
\author[2]{A.~Mazzolari}
\author[2]{G.~Patern\'{o}}
\author[2,3]{J.~Reyes~Garrido}
\author[2]{M.~Romagnoni}
\author[2]{A.~Sytov}
\author[4]{A.~V.~Korol}
\author[4]{A.~V.~Solov'yov}

\cortext[cor1]{Corresponding author}
\cortext[cor2]{Corresponding author}

\address[1]{Department of Physics and Earth Science, University of Ferrara,
            Via Saragat 1, 44122 Ferrara, Italy}

\address[2]{INFN, Ferrara Section,
            Via Saragat 1, 44122 Ferrara, Italy}

\address[3]{Department of Physics, Sapienza University of Rome,
            Piazzale Aldo Moro 5, 00185 Rome, Italy}

\address[4]{MBN Research Center,
            Altenh\"{o}ferallee 3, 60438 Frankfurt am Main, Germany}

\begin{abstract}
The development of compact and intense $\gamma$-ray sources in the MeV energy
range remains a significant frontier in radiation physics, with profound
implications for nuclear physics, medicine and applied science.
In this work, we present a comprehensive numerical investigation of the photon
emission probability and brilliance of a Crystalline Undulator (CU) based on a
periodically bent Si(110) crystal.
Our approach integrates realistic deformation profiles obtained via Finite Element
Method simulations of a realistic sample, where bending is induced by patterned
$\text{Si}_3\text{N}_4$ surface stressors.
Relativistic molecular dynamics simulations, performed using the
\textsc{MBN Explorer} software package, consider a 10~GeV positron beam consistent
with the foreseen FACET-II facility specifications.
Our results reveal a distinct undulator radiation peak in the 1.6--2.1~MeV
range, well-separated from the broader channeling radiation background.
We show that for an optimal aperture angle of $1/2\gamma$, the source reaches
a maximum peak brilliance of about
$5\times10^{22}$~photons/s/mm$^2$/mrad$^2$/0.1\%~BW.
This performance is highly competitive with large-scale Gamma-Beam Systems and
exceeds that of Inverse-Compton Scattering sources, confirming the potential of
crystalline undulators as high-brilliance, compact
light sources for the MeV domain.
\end{abstract}




\end{frontmatter}

\section{Introduction}\label{sec1}

The development of high-intensity light sources (LSs) capable of reaching photon energies well
above 10~keV remains a central challenge in modern accelerator and radiation physics.

Extending high brilliance operation toward the MeV domain would enable a wide range of
applications in fundamental science, nuclear physics and photo-nuclear studies that require hard
X-ray and $\gamma$-ray beams \cite{korol2020crystal, zhu-2020, Howell_2022}. Existing X-ray free
electron lasers (XFELs) routinely operate in the \AA ngstr\"{o}m regime, while synchrotron
facilities can reach hard X-rays at the price of a significant reduction in the brightness
\cite{doerr-2018, seddon-2017, milne-2017, bostedt-2016, couprie-2014, yabashi-2017}. Pushing
coherent emission to photon energies in the MeV to GeV range therefore calls for new physical
mechanisms and compact source concepts.

An innovative approach to overcome these limitations involves the generation of $\gamma$-rays
through inverse Compton scattering (ICS). ICS sources can in principle reach photon energies up to
the GeV range \cite{muramatsu-2014}, though their operation imposes demanding infrastructural and
technological requirements due to the need for integrated particle accelerators and laser systems.
In the hard $\gamma$-ray domain (above ${\sim}1$~MeV), the photon fluxes achievable at ICS
facilities are typically much lower than those of X-ray ICS sources, which can reach
$10^{11}$--$10^{13}$~photons/s \cite{petrillo-2023}; a compiled overview of $\gamma$-ICS
facilities and their performance can be found in \cite{sushko-2024}.

In recent years \cite{korol2020crystal, sushko-2024, sushko-2022} significant efforts of the
research and technological communities have been devoted to the design and practical realization of
novel gamma-ray crystal-based light sources (CLS), which exploit oriented crystals of different
geometries, linear, bent, or periodically bent, exposed to ultrarelativistic beams of electrons and
positrons. The practical realization of CLSs is currently being pursued within the Horizon Europe
EIC-Pathfinder project TECHNO-CLS \cite{techno-cls}.

A promising alternative is provided by crystalline undulators (CUs). A CU is a device in which
relativistic particles undergo channeling within a crystal that features periodically bent
crystallographic planes~\cite{korol-2014}. This periodic trajectory induces the emission of
coherent undulator radiation, effectively mimicking the operation of a conventional magnetic
undulator but on a much smaller scale, utilizing the extremely high electrostatic fields of the
lattice \cite{korol-1999}.

Periodic bending can be achieved through several other techniques, including mechanical grooving
\cite{camattari-2019}, pulsed laser melting \cite{di-russo-2022}, ion implantation
\cite{bellucci-2015}, graded composition layers \cite{avakian-2002}, acoustic waves
\cite{kaleris-2025}, boron-doped diamond superlattices \cite{thi-2017} or the use of surface
stressors \cite{guidi-2007, guidi-2011}. Within this framework, positrons are particularly
suitable for radiation generation. While electrons are attracted to the high-density atomic nuclei,
leading to higher scattering rates and dechanneling, positrons experience a nearly harmonic
interplanar potential. This ensures more stable trajectories and significantly higher channeling
efficiency.

In this work, we present accurate predictions for the spectral distribution and brilliance of a
crystalline undulator~\cite{korol-2014}, derived from all-atom molecular dynamics simulations of
relativistic particle channeling and radiation processes. Specifically, we numerically assessed the
photon emission from a periodically bent Si(110) crystal, whose deformation profile is induced by
Si$_3$N$_4$ surface stressors~\cite{malagutti-2025}. As a representative example, we adopt the
beam parameters of the 10~GeV/c positron beam proposed for the Stanford FACET-II facility
\cite{yakimenko-2019}. These parameters are compatible with beam characteristics available at
current positron sources and with those foreseen at other facilities in the near future.

This investigation represents a significant advancement over previous theoretical studies, which
typically relied on idealized sinusoidal bending models \cite{sushko-2022}. The radiation spectrum
and brilliance are computed starting from the realistic deformation profiles of a
laboratory-manufactured sample, simulated via Finite Element Method (FEM) analysis. This approach
provides a robust validation of the CU concept, demonstrating that even with the inclusion of
realistic structural constraints and edge effects, it is feasible to generate radiation in the
photon energy range $\hbar\omega\gtrsim1$~MeV with a brilliance exceeding that of ICS sources
\cite{Howell_2022, wu-2006, rehman-2017, sei-2020}.

Numerical modeling of channeling phenomena was performed using the \textsc{MBN Explorer} package
\cite{solovyov-2012, sushko-2013} and the supplementary multi-task toolkit \textsc{MBN Studio}
\cite{sushko-2019}, allowing for an analysis that extends beyond the standard continuous potential
framework. Specifically, the relativistic dynamics module within \textsc{MBN Explorer}
\cite{sushko-2013} enables the simulation of projectile motion coupled with a dynamic description
of the environment. A distinct advantage of this algorithm is its ability to account for the
interaction of projectiles with all constituent atoms of the medium, thereby avoiding simplifying
model assumptions and facilitating rigorous simulations of crystalline structures through a variety
of interatomic potentials. A comprehensive overview of channeling and radiation results in oriented
linear, bent, and periodically bent crystals obtained via this computational approach can be found
in \cite{korol2020crystal,korol-2014, korol-2021, solovyov-2017}.

\section{Simulation of the Crystalline Undulator via FEM}

\begin{figure}[ht]
    \centering
    \includegraphics[width=0.9\linewidth]{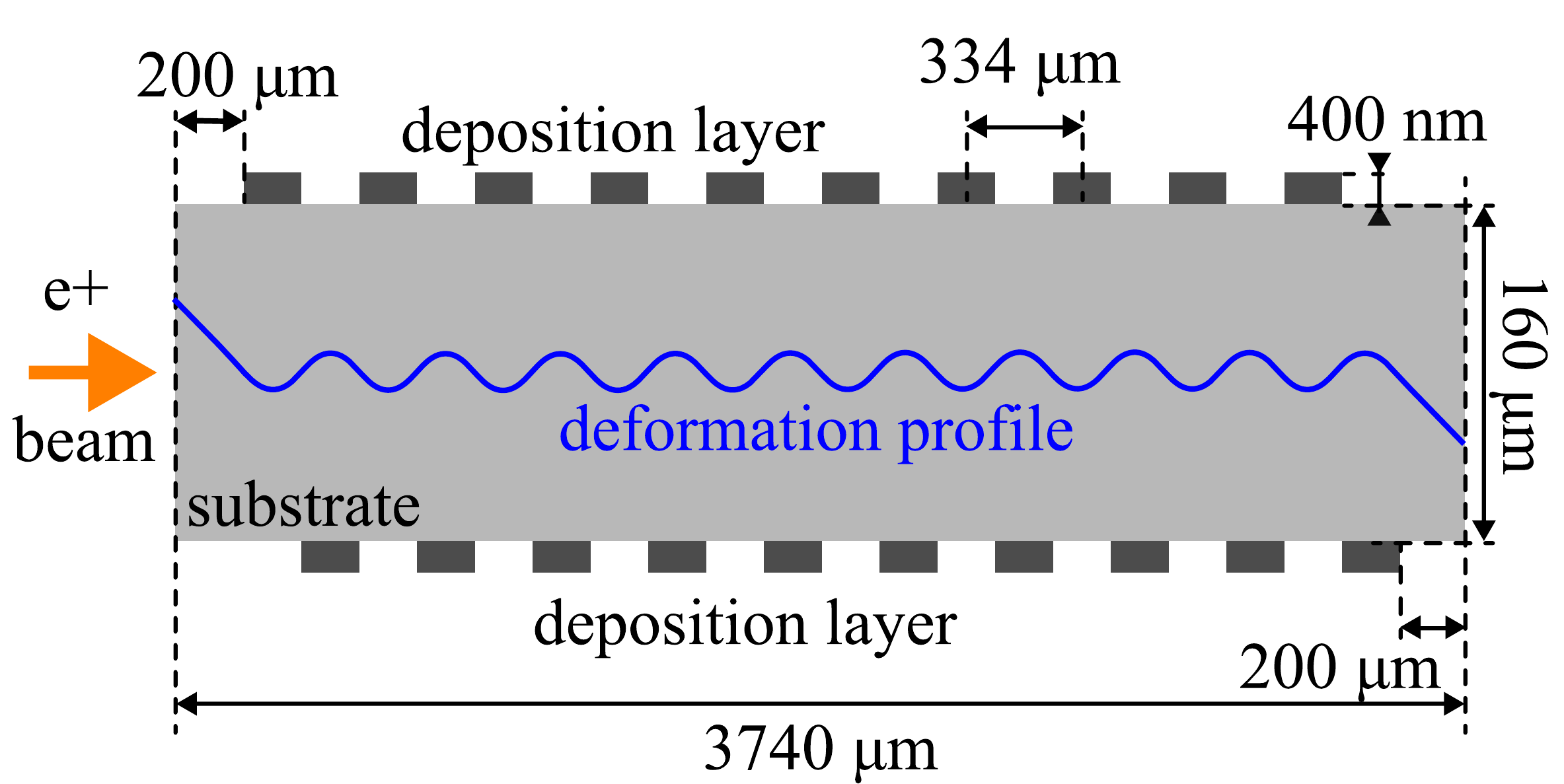}
    \caption{Schematic representation of the simulated crystal geometry. The silicon substrate
             is bent by the deposition of periodic $\text{Si}_3\text{N}_4$ strips on both
             surfaces.}
    \label{fig:sample}
\end{figure}

The CU investigated in this work relies on the periodic deformation of a silicon crystal, achieved
through the deposition of tensile-strained silicon nitride ($\text{Si}_3\text{N}_4$) layers. To
determine the exact deformation profile utilized for the channeling simulations, a static
structural analysis was performed using the Finite Element Method (FEM) via the ANSYS software
suite~\cite{malagutti-2025}.

The simulated geometry, illustrated in Fig.~\ref{fig:sample}, consists of a $160~\mu$m-thick
dislocation-free silicon substrate oriented along the (110) crystallographic plane. The substrate
is coated on both surfaces with a $400$~nm-thick $\text{Si}_3\text{N}_4$ film deposited via
Low-Pressure Chemical Vapor Deposition (LPCVD)~\cite{malagutti-2025}. To induce the periodic
bending moment required for the undulator effect, the nitride layer is patterned into rectangular
strips arranged with a periodicity of $\lambda_{\mathrm{CU}}=334~\mu$m for 10 periods. A phase
shift of $\lambda_{\mathrm{CU}}/2$ is introduced between the patterns on the front and back
surfaces to generate a symmetric sinusoidal deformation profile along the $3740~\mu$m length of
the device.

To accurately reproduce the intrinsic tensile stress generated during the LPCVD process, an
equivalent thermal load method was implemented as detailed in Ref.~\cite{malagutti-2025}. The
stress field was modeled by imposing a fictitious temperature difference derived from the mismatch
in thermal expansion coefficients between the silicon nitride film and the silicon substrate. The
mechanical behavior of the substrate was defined using the anisotropic elasticity matrix for cubic
crystals, while the amorphous nitride film was treated as an isotropic material.

\begin{figure}[ht]
    \centering
    \includegraphics[width=0.9\linewidth]{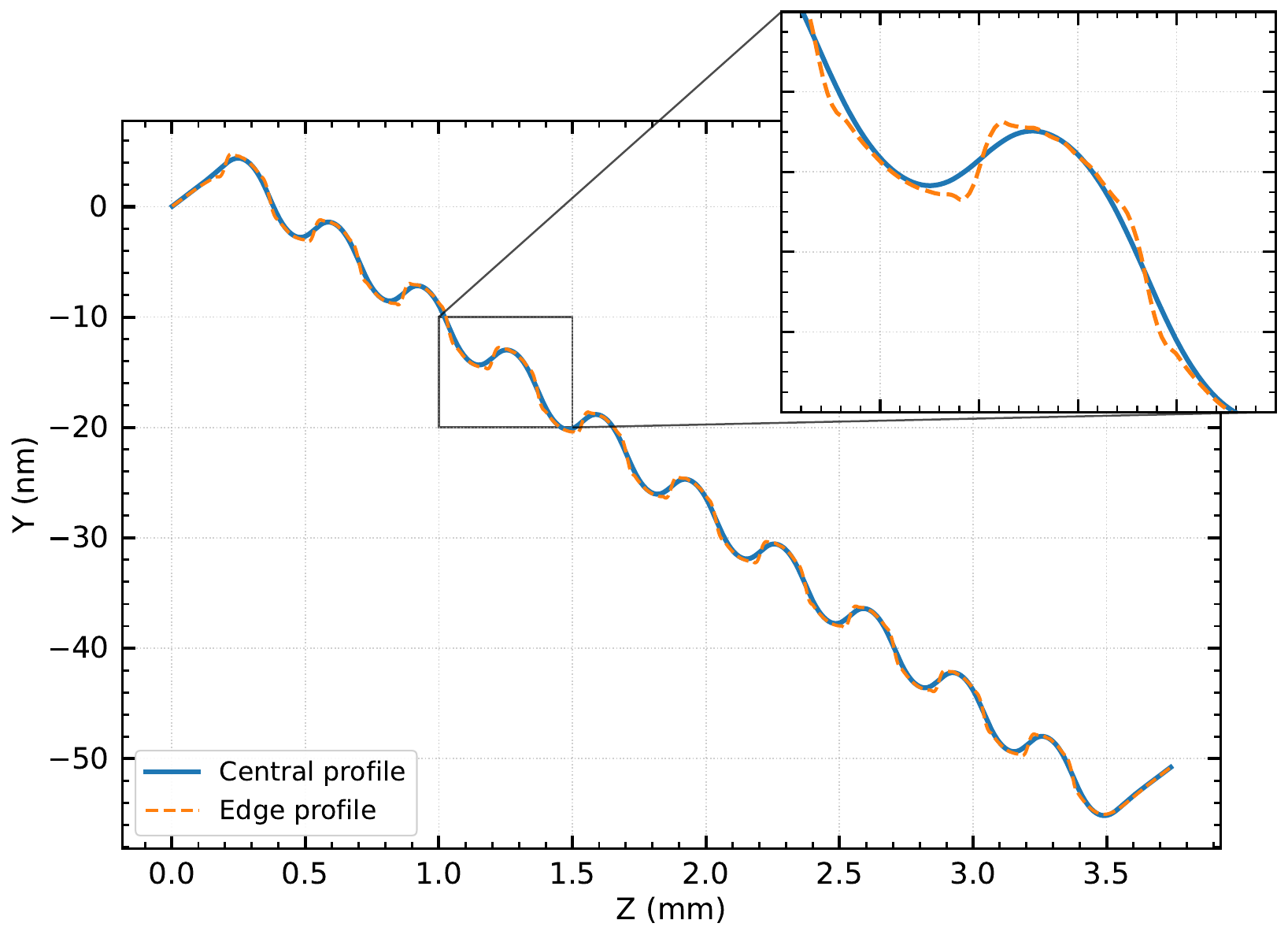}
    \caption{Deformation profiles at depths of $80~\mu$m (Central) and $10~\mu$m (Edge) from the
             surface. The inset highlights edge irregularities due to stress concentration near
             the strips.}
    \label{fig:Fig2}
\end{figure}

The resulting deformation profiles are presented in Fig.~\ref{fig:Fig2}. The static structural
analysis reveals a highly stable deformation field across the bulk of the sample. Specifically,
the profile extracted at the mid-plane ($80~\mu$m depth) exhibits a smooth, quasi-sinusoidal shape
strongly dominated by the fundamental harmonic. Notably, the FEM data demonstrate that this ideal
harmonic behavior remains remarkably uniform across a broad central region of approximately
$100~\mu$m in width.

As expected, the profiles extracted closer to the crystal surfaces (e.g., $10~\mu$m from the
edge) display localized irregularities. These deviations are strictly confined to the boundary
regions and arise from the stress concentration near the discontinuous edges of the
$\text{Si}_3\text{N}_4$ patterned strips.

Crucially, according to the nominal FACET-II beam parameters detailed in Sec.~\ref{sec2-Defl},
the transverse size of the incident positron beam is $\sigma_{x,y}\simeq23$ to $25~\mu$m. This
implies that the vast majority of the particle population is deeply confined within the highly
uniform $100~\mu$m central bulk of the crystal, intersecting almost exclusively the ideal
quasi-sinusoidal portion of the lattice. Therefore, adopting the optimized mid-plane profile as a
uniform characteristic of the active crystal volume for the relativistic molecular dynamics
simulations is not merely a computational simplification, but a physically robust choice. The beam
effectively bypasses the surface irregularities, rendering edge effects entirely negligible for
both the particle dynamics and the overall radiation generation process.

\section{Simulation Methodology and Spectral Analysis}\label{sec2}

The numerical simulations were performed using \textsc{MBN Explorer}, which computes, for each
individual trajectory $j$, the spectral distribution of the energy radiated into a cone of opening
angle $\vartheta_0$: $\text{d}E_j(\vartheta\leq\vartheta_0)/\text{d}(\hbar\omega)$.
The full spectral distribution is then obtained by averaging over
$N_\mathrm{tot}\approx10^3$ simulated trajectories:

\begin{equation}
\frac{\text{d}E(\vartheta\leq\vartheta_0)}{\text{d}(\hbar\omega)} =
N_\mathrm{tot}^{-1}\sum_{j=1}^{N_\mathrm{tot}}
\frac{\text{d}E_j(\vartheta\leq\vartheta_0)}{\text{d}(\hbar\omega)}
\label{eq:avg_spectrum}
\end{equation}

The initial conditions at the crystal entrance were generated using normal distributions with
deviations $\sigma_{x,y}$ and $\sigma_{\phi_{x,y}}$ corresponding to the FACET-II beam
parameters.

For visualization purposes, the averaged energy spectrum is converted into the photon emission
probability per unit energy per positron:
\begin{equation}
\frac{\text{d}n_\gamma(\vartheta\leq\vartheta_0)}{\text{d}(\hbar\omega)} =
\frac{1}{\langle\hbar\omega\rangle}
\frac{\text{d}E(\vartheta\leq\vartheta_0)}{\text{d}(\hbar\omega)}
\label{eq:photon_prob}
\end{equation}
where $\langle\hbar\omega\rangle$ is the central photon energy of each bin.

The spectral brilliance $B_\omega$ of the source is defined as the number of photons
$\Delta n_\gamma$ within the frequency interval
$[\omega-\Delta\omega/2,\,\omega+\Delta\omega/2]$ emitted into the cone angle $\Delta\Omega$,
per unit time, unit source area, unit solid angle, and per bandwidth $\Delta\omega/\omega$.
Following the methodology of Ref.~\cite{sushko-2022}, it can be expressed as
(see, e.g.,~\cite{schmuser-2008}):

\begin{equation}
B_{\omega} = \frac{\text{d}E(\vartheta\leq\vartheta_0)}{\text{d}(\hbar\omega)}
\frac{1.58\times10^{14}\,I}{\Sigma_x\Sigma_y}
\label{eq:brilliance}
\end{equation}
where $E_\gamma\equiv\hbar\omega$ is the photon energy;
$1.58\times10^{14}=10^{-3}/[e(2\pi)^2]$ with $e$ the elementary charge; $I$ is the beam
current (in A); and $\Sigma_{x,y}=(\sigma^2+\sigma_{x,y}^2)^{1/2}(\phi^2+\sigma_{\phi_{x,y}}^2)^{1/2}$
is the total transverse emittance of the photon source~\cite{schmuser-2008}, with $\sigma$ in mm
and $\phi$ in mrad, yielding $B_\omega$ in
$[\text{photons/s/mrad}^2/\text{mm}^2/0.1\%~\text{BW}]$.

\section{Radiation Generation and Brilliance}
\label{sec2-Defl}

For this study, the simulated radiation corresponds to the beam parameters foreseen for the
10~GeV positron line of FACET-II, as reported in the facility's Technical Design Report (TDR)
\cite{yakimenko-2019}. At this energy, the Lorentz factor is
$\gamma=E/mc^2\simeq1.96\times10^4$, giving $1/\gamma\simeq50~\mu$rad, while for the Si(110)
planes the Lindhard critical angle is $\vartheta_C\approx60~\mu$rad.

\begin{figure*}[ht]
    \centering
    \includegraphics[width=0.95\linewidth]{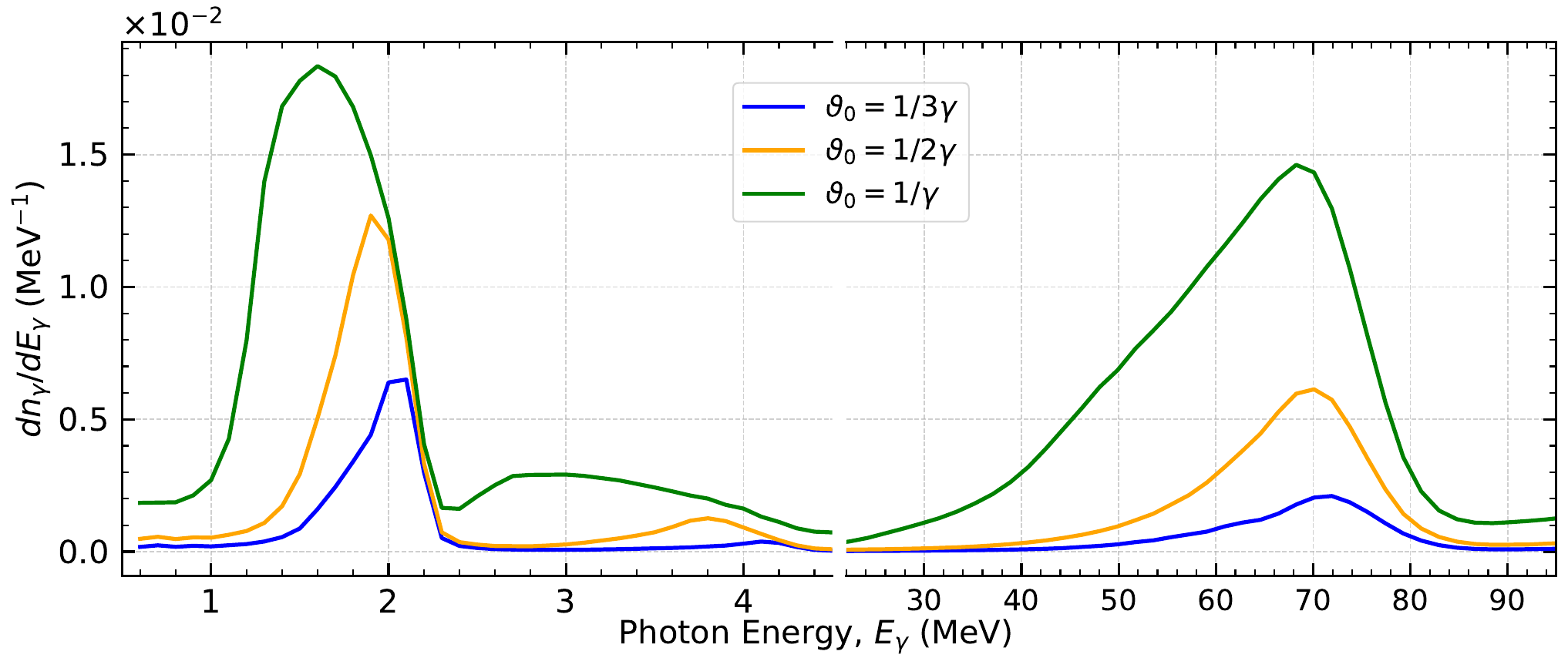}
    \caption{Simulated photon emission probability for the different opening angles $\vartheta_0$:
             $1/(3\gamma)$ (blue), $1/(2\gamma)$ (orange), and $1/\gamma$ (green).}
    \label{fig:undurad_and_ChR}
\end{figure*}

Consistent with the nominal normalized emittances of FACET-II
($\varepsilon_{n,x}\simeq10.7~\mu\mathrm{m\,rad}$,
$\varepsilon_{n,y}\simeq13~\mu\mathrm{m\,rad}$), the simulations adopted beam transverse sizes of
$\sigma_x\simeq23~\mu\mathrm{m}$, $\sigma_y\simeq25~\mu\mathrm{m}$ and divergences of
$\sigma'_x\simeq23~\mu\mathrm{rad}$, $\sigma'_y\simeq25~\mu\mathrm{rad}$. These divergence
values ensure the beam is well within the channeling acceptance limits
($\sigma'_{x,y}<\vartheta_C$).

To reproduce the optimal experimental alignment, the simulated beam direction was adjusted by an
angular offset of approximately $18.3~\mu$rad. This value corrects for the average tilt of the
crystal mid-plane profile derived from the FEM analysis.

The subsequent step involved the calculation of the emitted radiation spectra.
Figure~\ref{fig:undurad_and_ChR} shows the photon emission probability
$dn_\gamma/d(\hbar\omega)$ as defined in Eq.~(\ref{eq:photon_prob}), computed for three opening
angles $\vartheta_0$: $1/(3\gamma)=17.8~\mu$rad, $1/(2\gamma)=26.5~\mu$rad, and
$1/\gamma=51.1~\mu$rad. As seen in Fig.~\ref{fig:undurad_and_ChR}, the spectra are composed of
two major features. The first is the high-energy component, corresponding to channeling radiation,
with a peak located around 70~MeV and a sharp drop beyond 80~MeV. The position of this peak is
in excellent agreement with theoretical expectations.

The second, and more relevant contribution in this study, is the undulator radiation peak. For
all observation angles, this peak appears more intense than the channeling radiation and is located
around 1.6--2.1~MeV.

In terms of spectral characteristics, this feature is significantly distinct from the broader
channeling background. Quantitatively, the undulator peak exhibits a Full Width at Half Maximum
(FWHM) of approximately 0.5~MeV. This corresponds to a natural bandwidth of ${\sim}25\%$, a
broadening consistent with the intrinsic properties of a short undulator ($N=10$) and the effects
of multiple scattering. Nevertheless, the spectral density remains high enough to provide a
quasi-monochromatic source in the MeV range.

A zoomed view of the 0.5--6~MeV region is provided in Fig.~\ref{fig:undurad_and_ChR},
highlighting the dependence of the peak position on the observation angle. In the analysis that
follows, the peak brilliance is normalized to the standard 0.1\% bandwidth (BW) to facilitate
direct comparison with other light sources~\cite{sushko-2024, sushko-2022}, while acknowledging
the larger physical bandwidth of the emitted radiation.

\begin{table*}[!ht]
\centering
\renewcommand{\arraystretch}{1.15}
\caption{Energy at the peak and the maximum peak brilliance for the different $\vartheta_0$.}
\label{tab:tabCU}
\begin{tabular}{ccc}
\hline
$\vartheta_0$ [$\mu$rad] & $E_{\text{peak}}$ [MeV] &
  $B_{\text{peak}}^{\text{max}}$ [ph/s/mm$^2$/mrad$^2$/0.1\%~BW] \\
\hline
17.8 ($1/3\gamma$) & 2.1 & $3.9\times10^{22}$ \\
25.6 ($1/2\gamma$) & 1.9 & $5.3\times10^{22}$ \\
51.1 ($1/\gamma$)  & 1.6 & $3.1\times10^{22}$ \\
\hline
\end{tabular}
\end{table*}

\begin{figure}[htbp]
    \centering
    \includegraphics[width=0.9\linewidth]{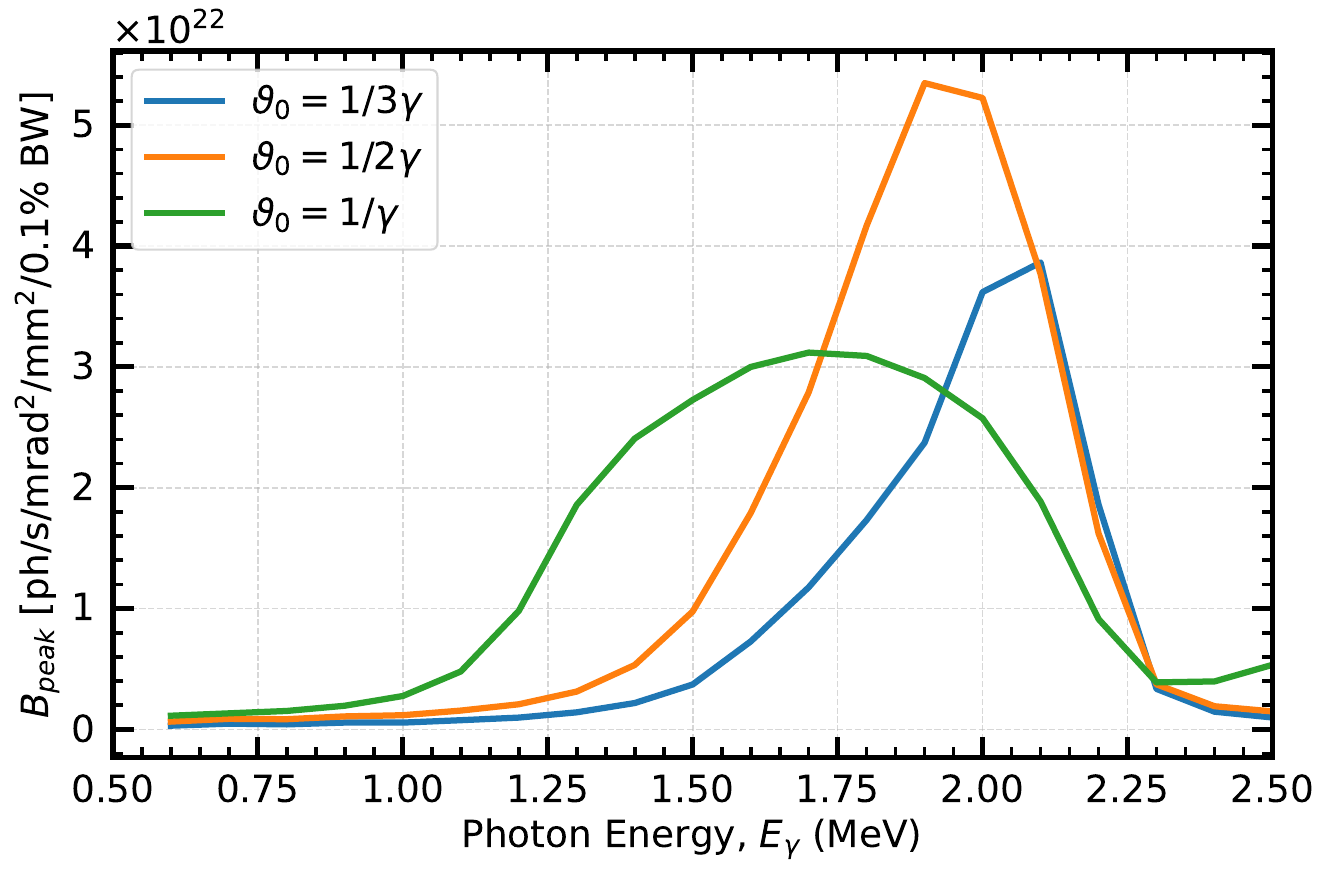}
    \caption{Peak brilliance as a function of the photon energy for different aperture angles
             $\vartheta_0$: $1/(3\gamma)$ (blue), $1/(2\gamma)$ (orange), and $1/\gamma$
             (green).}
    \label{fig:BpeakVsE}
\end{figure}

To quantify the source performance, the peak brilliance $B_{\text{peak}}$ was evaluated from
these spectra using the methodology described in Sec.~\ref{sec2}. We assumed a peak current
$I_{\text{peak}}=6$~kA, consistent with the nominal design parameters reported for
high-compression operations at FACET-II~\cite{yakimenko-2019}. The brilliance curves in the
0.5--2.5~MeV region are presented in Fig.~\ref{fig:BpeakVsE}.

The analysis demonstrates that the effective brilliance is sensitive to the aperture angle
$\vartheta_0$. Our results indicate that $\vartheta_0\approx1/2\gamma$ represents the optimal
trade-off point, maximizing the peak brilliance. At smaller angles ($1/3\gamma$), the loss of
photon flux outweighs the gain in phase-space reduction, while at larger angles ($1/\gamma$), the
phase-space volume grows too large, diluting the brilliance. Table~\ref{tab:tabCU} summarizes,
for each aperture angle, the peak photon energy $E_\mathrm{peak}$ of the undulator radiation and
the corresponding maximum peak brilliance $B_\mathrm{peak}^\mathrm{max}$, evaluated from
Eq.~(\ref{eq:brilliance}). The highest value,
$5.3\times10^{22}$~ph/s/mm$^2$/mrad$^2$/0.1\%~BW is achieved for this optimal observation angle
at a photon energy of 1.9~MeV.

These results confirm that the CU-LS can provide high-intensity, quasi-monochromatic radiation in
the MeV range, significantly exceeding the performance of conventional Laser-Compton Scattering
sources in the same energy domain.

\section{Conclusions}\label{sec:conclusions}

\begin{table*}[ht]
\centering
\caption{Comparison of the maximum peak brilliance across different photon sources. Brilliance is
         expressed in units of photons/s/mm$^2$/mrad$^2$/0.1\% BW.}
\label{tab:comparison}
\begin{tabular}{lcc}
\hline
Facility & Energy Range & Peak Brilliance \\
\hline
European XFEL \cite{XFEL-2025}  & 10--20 keV  & ${\sim}10^{33}$ \\
ESRF-EBS \cite{ESRF-2025}       & 30--100 keV & $10^{20}$--$10^{21}$ \\
ELI-NP \cite{ur-2015}           & 0.2--20 MeV & $10^{20}$--$10^{23}$ \\
CU (This work)                  & 1.6--2.1 MeV & $5.3\times10^{22}$ \\
\hline
\end{tabular}
\end{table*}

The radiation spectra, computed with the Baier--Katkov quasiclassical formalism in
\textsc{MBN Explorer}, exhibit a narrow undulator peak in the 1.6--2.1~MeV range, superimposed
on the bremsstrahlung background. These results confirm that the periodically bent Si(110) crystal
investigated here can act as a compact, quasi-monochromatic $\gamma$-ray source.

To evaluate the performance of the crystalline undulator (CU), its peak brilliance is compared
with other state-of-the-art facilities in Table~\ref{tab:comparison}.

As shown in Table~\ref{tab:comparison}, although the peak brilliance of the CU is lower than that
of large-scale X-ray free electron lasers, it remains highly competitive within the high-energy
photon domain. In the MeV energy range, the proposed source performs similarly to the ELI-NP
$\gamma$-beam system. Furthermore, compared to storage-ring sources like ESRF-EBS, the CU concept
offers a gain of two to three orders of magnitude in brilliance while significantly extending the
accessible photon energy. The source intensity can be further optimized by extending the crystal
length, provided the characteristic dechanneling length of the positrons is not exceeded
\cite{korol2020crystal, korol-2014}.

While storage rings provide superior average brilliance due to high repetition rates, the CU
driven by a linac like FACET-II excels in peak brightness. This makes it ideal for single-shot
experiments requiring high-intensity pulses in the MeV domain. The development of high-energy
electron facilities ($>$1~GeV) is currently driven by a variety of research goals beyond light
sources, including plasma wakefield acceleration (PWFA), dark matter searches, and nuclear physics
experiments. Within this landscape, future high-brilliance machines for positrons are also being
planned. The integration of crystalline undulators into such facilities could provide a unique
pathway to generate high-intensity photon beams for advanced nuclear spectroscopy and industrial
radiography. These results demonstrate that CUs are a viable and compact alternative to
conventional technologies for MeV photon production.

\section*{Acknowledgements}

This work was supported by the European Commission through the H2020-MSCA-RISE N-LIGHT
(Grant Agreement No.\ 872196) and the EIC-PATH\-FINDER-OPEN TECHNO-CLS project
(Grant Agreement No.\ 101046458).




\section*{Declaration of competing interests}

The authors declare that they have no known competing financial interests or personal
relationships that could have appeared to influence the work reported in this paper.


\section*{Data availability}

The simulation data that support the findings of this study are available from the
corresponding author upon reasonable request.

\bibliographystyle{elsarticle-num}
\bibliography{sn-bibliography}

\end{document}